# Electronic properties and phase transition in Kagome metal, Yb$_{0.5}$Co$_3$Ge$_3$

Yaojia Wang[1,2], Xiaoping Wang[4], Gregory T. McCandless[3], Kulatheepan Thanabalasingam[3], Heng Wu[1,2], Damian Bouwmeester[1,2], Herra van der Zant[1,2], Mazhar N. Ali[1,2]*, Julia Y. Chan[3]*

1) Kavli Institute of Nanoscience, Delft University of Technology, Delft, the Netherlands
2) Department of Quantum Nanoscience, Faculty of Applied Sciences, Delft University of Technology
3) Department of Chemistry and Biochemistry, Baylor University, Waco, TX, 76798, USA
4) Neutron Scattering Division, Oak Ridge National Laboratory, Oak Ridge, Tennessee 37831, USA
*Corresponding author: M.N.Ali@tudelft.nl, Julia_Chan@baylor.edu

The Kagome lattice is an important fundamental structure in condensed matter physics for investigating the interplay of electron correlation, topology, and frustrated magnetism. Recent work on Kagome metals in the AV$_3$Sb$_5$ (A = K, Rb, Cs) family, has shown a multitude of correlation-driven distortions, including symmetry breaking charge density waves and nematic superconductivity at low temperatures. Here we study the new Kagome metal Yb$_{0.5}$Co$_3$Ge$_3$ and find a temperature-dependent kink in the resistivity that is highly similar to the AV$_3$Sb$_5$ behavior and is commensurate with an in-plane structural distortion of the Co Kagome lattice along with a doubling of the *c*-axis. The space group is found to lower from *P*6/*mmm* to *P*6$_3$/*m* below the transition temperature, breaking the in-plane mirror planes and C$_6$ rotation, while gaining a screw axis along the *c*-direction. At very low temperatures, anisotropic negative magnetoresistance is observed, which may be related to anisotropic magnetism. This raises questions about the types of the distortions in Kagome nets and their resulting physical properties including superconductivity and magnetism.

## Introduction

Materials containing 2D Kagome lattices, also known as trihexagonal tiling, are of extreme interest in condensed matter physics today. The Kagome lattice is made up of hexagons that are surrounded on their edges by equilateral triangles. Analogous to the honeycomb lattice, the Kagome lattice is an important structure for realizing a quantum spin liquid state due to its inherent geometric frustration; some of the leading spin liquid candidates are Kagome insulators [1,2]. Recently, however, Kagome metals have gained attention due to the realization of both topological Dirac electrons as well as flat bands and van Hove singularities resulting in strong electron correlation [3–7]. Materials in the AV$_3$Sb$_5$ (A = K, Cs, Rb) family [8–11] have shown both topological bands with low effective mass, large anomalous Hall effect [12,13], as well as a cascade of charge density wave (CDW) [14–22] orderings and even nematic superconductivity [23–29] as temperature is lowered.

The intertwining of orders seen in Kagome metals is a complicated interplay of electronic correlation, topology, and magnetism resulting in nearby ground states that are accessible by temperature control. While some phases (AV$_3$Sb$_5$) have shown electronic correlation related charge orders and topology, others have shown topology and magnetism [30–38]. Studies are ongoing on Kagome systems which show correlation and magnetism, where charge/spin orders intertwine with magnetic orders of a spin sublattice. It has been shown that the structural distortions of the Kagome lattice and symmetry breaking are closely related to the appearance of new orders, including the symmetry breaking CDW order observed in AV$_3$Sb$_5$ compounds, and antiferromagnetic FeGe [39]. How the various orders work together, or compete with each other, and what effect that has on structural distortions in the Kagome net and the resulting physical properties, remains an active area of investigation.

In this paper, we report the presence of a kink in the resistivity at 95 K, below which there is a structural phase transition resulting in the distortion of the Co Kagome net, in the new Kagome metal $Yb_{0.5}Co_3Ge_3$. Through single crystal X-ray diffraction structure determination, the low temperature structure was found to break $C_6$ rotation symmetry and out-of-plane mirror planes, lowering symmetry from the *P6/mmm* space group to *P6₃/m*. In addition, a small upturn is evident in the resistivity below 18 K, commensurate with a previously seen transition in the magnetization. Below this temperature, anisotropic negative magnetoresistance is observed, in contrast with the positive magnetoresistance seen in other Kagome metals like $AV_3Sb_5$. $Yb_{0.5}Co_3Ge_3$ is a good platform for studying the effect of correlation and magnetism on the electronic properties in a Kagome system and adds to the ongoing effort to understand the phase space of Kagome metals.

## Methods

*Synthesis and property measurement.* Single crystals of $Yb_{0.5}Co_3Ge_3$ were grown as reported previously [40]. Elements were weighed out into a Canfield crucible set [41] with the molar ratio of 3:2:7:52 Yb: Co: Ge: Sn, respectively, and sealed in a fused silica tube filled with argon gas ~1/3 atm pressure. The sealed ampoule was then heated in a furnace to 1175 °C at 100 °C/h and dwelled for 24 hours before cooling the ampoule down to 815 °C at a rate of 3 °C/h. The ampoule was then removed from the furnace, inverted, and centrifuged to remove the excess flux from the crystals. The needle-shaped crystals were then etched with dilute HCl followed by dilute $HNO_3$ etching to remove residual flux from the crystal surface. Silver paste was used to make contact with $Yb_{0.5}Co_3Ge_3$ crystals and the electrical properties were measured in a Quantum Design Physical Property Measurement System (PPMS) using the four-probe method with an a.c. current applied along the *c*-axis.

*Single Crystal X-ray Diffraction.* For comparison, the crystal structure of $Yb_{0.5}Co_3Ge_3$ collected at room temperature [40] is compared to the low temperature model. The 50 K single crystal X-ray diffraction data were collected using a Rigaku XtaLAB AFC12(RCD3) diffractometer equipped with graphite monochromated Mo Kα radiation (λ = 0.71073 Å), HyPix-6000He area detector, and the Rigaku Oxford Diffraction CrysAlisPro software. Sample temperature was controlled with the dual flow nitrogen and helium gas cooler N-Helix by Oxford Cryosystems. The numerical absorption correction was completed based on gaussian integration over a multifaceted crystal model. The empirical absorption correction was carried out using spherical harmonics implemented in SCALE3 ABSPACK scaling algorithm in CrysAlis PRO 1.171.41.123a (Rigaku Oxford diffraction 2022). The 50 K data was modeled in Jana 2006 software [42] using the room temperature structure as the preliminary structural model.

## Results and discussion

$Yb_{0.5}Co_3Ge_3$ crystallizes in the hexagonal *P6/mmm* space group at room temperature, which adopts a hybrid structure of the $YCo_6Ge_6$ and CoSn prototypes [43–45] with a Co Kagome lattice [40]; the crystal structure is shown in Fig.1a. Single crystals of $Yb_{0.5}Co_3Ge_3$ present a rod-like shape with the long axis being the *c*-axis [40]. Fig. 1b shows the temperature dependent resistivity $\rho(T)$ curve, which presents metallic behavior with reducing temperature. Several points of interest are observed on $\rho(T)$ curve; first is an evident kink near $T^* \sim$ 95 K with a clear transition on the $d\rho(T)/dT$ vs $T$ curve (lower inset Fig. 1b). This is found to correspond to a structural phase transition, which will be discussed in detail below. Below ~ 18 K, a weak up-turn of the resistivity is observed (also shown by the change in sign of the $d\rho(T)/dT$ vs $T$ curve in the inset), followed by a superconductor-like transition ~3.6 K (upper inset Fig. 1b). Above the up-turn of in the resistivity, the low-temperature $\rho(T)$ data (23 K – 60 K) is well fitted by the equation $\rho(T)=\rho(0)+AT^n$ (black line in Fig.1b), with $n$ = 2.04, $\rho(0)\approx$14.7 $\mu\Omega$ cm and $A\approx$0.00127 $\mu\Omega$ cm K$^{-2}$, suggesting electron-electron interactions dominate electronic transport in this regime [33,46].

The magnetoresistance is measured to study the magnetic response at low temperature. Figure 1c shows the $\rho(H)$ data measured with magnetic field applied in the Kagome plane ($H \perp I$) and perpendicular to the Kagome plane ($H//I$) at 2 K. The rapid increase in $\rho(H)$ at very small field ($\mu_0H_c \sim$ 15 mT for $H \perp I$, and $\mu_0H_c \sim$ 20 mT for $H//I$) confirms the superconducting transition. Since the resistivity does not reach zero (only an ~30 % drop), and the superconducting signal is very close to the superconductivity of Sn ($T_c \sim$ 3.72 K, $\mu_0H_c \sim$ 30 mT), it is likely that the superconductivity is not intrinsic, but is rather coming from a Sn flux residual, although the sample was carefully centrifuged after growth and etched in dilute HCl and HNO$_3$ to remove the residual of Sn flux. More careful investigation to extremely low temperature is necessary to search for superconductivity in this material, which is beyond the scope of this work.

After breaking superconductivity, a negative magnetoresistance is observed to high field, which is stronger for the magnetic field applied along the $c$-axis ($H//I$) compared with applying magnetic field in Kagome plane ($H \perp I$). Kagome metals with weak magnetism typically present positive magnetoresistance, such as the AV$_3$Sb$_5$ family [12,13], and some CoSn-type materials [47]. Negative magnetoresistance has been seen in ferromagnetic Kagome materials, and was also proposed to occur from electron correlation induced ferromagnetic spin fluctuations [47,48]. For Yb$_{0.5}$Co$_3$Ge$_3$, an earlier study on the magnetic behavior indicated the compound has antiferromagnetic coupling without clear long range magnetic order [40]. In that study, below 15-20 K, a weak transition of magnetization was reported which was proposed to arise from spin canting or spin reorientation, and a larger magnetization was reported for out-of-plane applied magnetic field (along the $c$-axis) compared with in the Kagome plane [40]. The observed up-turn of resistivity (~18 K) in this work aligns very well with the temperature of the previously reported magnetization transition. Taken together, the anisotropic negative magnetoresistance shown in Fig. 1c is likely related to anisotropic magnetism in Yb$_{0.5}$Co$_3$Ge$_3$ at low temperature. Detailed study on the magnetic behavior of the phase in the future may reveal the origin of the transition and the negative magnetoresistance.

The main result of this work is the prominent transition near 95 K in the $\rho(T)$. Very similar transition features were observed in the $\rho(T)$ of some Kagome metals, such as the AV$_3$Sb$_5$ (A=K, Cs, Rb) family [8] and the antiferromagnetic Kagome metal FeGe [39], between 75-100 K. These transitions were found to be the result of superlattice formation from charge density wave ordering [15–18,39]. To reveal whether there was a corresponding structural transition in Yb$_{0.5}$Co$_3$Ge$_3$, we performed single crystal X-ray diffraction above and below the transition temperature. Figures 2a and 2b show the projection of the X-ray diffraction pattern along the $a$-axis ($b$-$c$ plane) and $c$-axis ($a$-$b$ plane) in reciprocal space, respectively. In the (0kl) plane ($b$-$c$ plane), a new set of diffraction peaks (marked by the orange box) can be seen in the 50 K pattern, located at half-integer spacing along the $l$ direction ($c$-axis in real space, Fig. 2a), compared with the diffraction pattern measured at room temperature. These new peaks arise from the formation of a superlattice with a doubled unit cell along the $c$-axis. In (hk0) plane ($a$-$b$ plane), a new set of diffraction peaks (marked by the yellow circles) is also observed at 50 K compared with the 297 K pattern (Fig. 2b). These peaks do not arise from a superlattice formation in the $a$-$b$ plane, but instead become visible at 50 K due to distortion of the Kagome lattice.

The lattice parameters of Yb$_{0.5}$Co$_3$Ge$_3$ obtained at 297 K and at 50 K from the single-crystal X-ray diffraction are provided in Table 1, and the corresponding crystal structure is shown in Figure 3. The crystal structure of Yb$_{0.5}$Co$_3$Ge$_3$ can be described as a stacking of alternating subunits – a subunit with a hexagonal array (or honeycomb arrangement) of Ge atoms that can be partially stuffed in-plane with Yb and the transition metal (Co) Kagome net that contains Ge atoms within the hexagons which can be displaced out-of-plane in response to partial occupation of the Yb above or below them (Fig. 3(a) and (b)). The spacing between the planes of these subunits along the $c$-axis is ~1.96 Å at room temperature and ~1.94 Å at 50 K, a relatively insignificant contraction along the $c$-axis. At room temperature, the tiling of hexagons and triangles in the Kagome net adopt ideal internal bond angles of 120° and 60°, respectively, with both types of geometric arrangements containing a uniform Co-Co interatomic distance of ~2.55 Å, being a perfect Kagome net, as shown in Fig. 3(c).

At low temperature, the crystal structure of $Yb_{0.5}Co_3Ge_3$ is best modeled in the $P6_3/m$ space group due to a geometric distortion within the plane of the Kagome net. This is not driven by either the Yb or Ge atoms, which retain their geometric relations very closely with the room temperature structure, but rather from the deviation of the Co atoms from the ideal positions of the Kagome lattice. The triangular arrangements remain unaltered (or temperature independent) in their interatomic distances, but are slightly rotated relative to each other (Fig. 3(d)), resulting in the hexagonal arrangements becoming distorted. The bond angles in the hexagons therefore deviate by ± ~7.5° from the ideal internal bond angles of 120° in an alternating fashion when comparing angles between the edges at neighboring vertices of the hexagons. This results in a deformation of the Kagome net; a twisting of the triangular units with respect to each other that breaks $C_6$ rotational and inversion symmetry, and results in the loss of all mirror planes parallel to the $c$-axis, subsequently reducing the space group to $P6_3/m$ from the original $P6/mmm$. Additionally, the geometric distortion of the Kagome subunit changes the long-range ordering along the $c$-axis. At room temperature, the unit cell is defined with the shorter $c$-axis (c ~ 3.91 Å) and one Kagome subunit located at $z = ½$. At $T = 50$ K, the long-range ordering along the $c$-axis requires a doubling (c ~ 7.78 Å) and the unit cell now contains two Kagome subunits located at $z = ¼$ and ¾ (Fig. 3(b) and 3(e)). These are related to each other by a sixfold screw axis along $c$, and an inversion center is located between the Kagome nets making the full structure centrosymmetric, even though a single distorted Kagome net breaks inversion symmetry (Fig. 3(d)). A similar distortion has been seen in the Kagome metals $MgCo_6Ge_6$ [49], $LaRu_3Si_2$ and $YRu_3Si_2$, some of them are superconductors but show no sign of magnetic or CDW transitions above $T_c$ [50–54].

In summary, we investigated the crystal structure and transport properties of the Kagome metal $Yb_{0.5}Co_3Ge_3$ and found a symmetry lowering structural phase transition around 95 K commensurate with a kink in the resistivity that is very similar to the CDW transition found in the $AV_3Sb_5$ and FeGe Kagome metals. Based on our measurements down to 2 K, we do not find intrinsic superconductivity, however we do observe anisotropic negative magnetoresistance is seen that correlates with previous study on a magnetic transition around 18 K. The 95 K transition distorts the Kagome net, keeping the triangular units consistent but rotating them slightly relative to each other, breaking $C_6$ rotation, out of plane mirror, and inversion symmetry. Since it is known that the structure of the Kagome lattice drives both flat bands, Dirac bands, and magnetic frustration in Kagome metals, resulting in both strong electron correlation, topological electrons, and complex magnetic states, which significantly influences the electronic properties, $Yb_{0.5}Co_3Ge_3$ and its distortion merits further theoretical and experimental investigation. Angle-resolved photoemission spectroscopy with associated band structure calculations can be performed to understand the band structure modifications associated with the phase transition and correlations in this material. Of particular importance is determining whether the structural transition is associated with formation of charge density wave order in analogy to other Kagome metals, ideally studied using scanning tunneling microscopy (STM). As recently discovered in the $AV_3Sb_5$ family, where multiple charge ordered phases were found as a function of decreasing temperature via STM but were hidden in resistive transport measurements, there may be further orderings present in $Yb_{0.5}Co_3Ge_3$ which remain to be found. Additionally, the magnetism in $Yb_{0.5}Co_3Ge_3$ at low temperature may also induce spin orders and muon spin relaxation measurements could elucidate this. Finally, since superconductivity has been observed in some other materials with distorted Kagome lattices as well as nematic superconductivity in the $AV_3Sb_5$ family, further studies based on chemical doping or high-pressure approaches at lower temperature can be used in Yb$_{0.5}$Co$_3$Ge$_3$ to probe this and its potential relation to the structural transition.

## Acknowledgements

Y.W. acknowledges the support from NWO Talent Program Veni financed by the Dutch Research Council (NWO), project No. VI.Veni.212.146. A portion of this research used resources at the Spallation Neutron Source, a Department of Energy Office of Science User Facility operated by the Oak Ridge National Laboratory. JYC acknowledges NSF-DMR 2209804 and Welch AT-2056-20210327 for partial support of this work. MNA acknowledges support from the Kavli Institute of Nanoscience Delft and the Faculty of Applied Sciences at TU Delft and the Max Planck Institute for Microstructure Physics Halle.

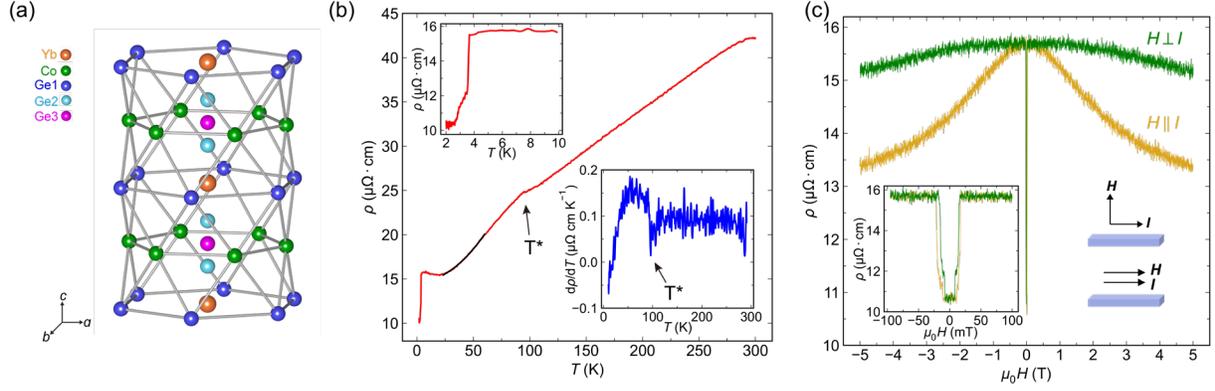

FIG. 1. Structure and properties of $Yb_{0.5}Co_3Ge_3$. (**a**). Crystal structure of $Yb_{0.5}Co_3Ge_3$ in the *P6/mmm* phase. (**b**). Temperature dependence of the resistivity with current applied along *c*-axis. The top inset is the $\rho$ vs $T$ curve at low temperature, the bottom inset is the $d\rho/dT$ vs $T$ curve. The transition near 95 K is marked by a black arrow. The black line on the $\rho$ vs $T$ curve in the main panel is a fit to $\rho(T)=\rho(0)+AT^n$. (**c**). Magnetic field dependence of the resistivity measured at 2 K. The inset on the left is the zoom-in of the superconductivity at small field. A schematic of the applied magnetic field direction for *H//I* and *H* ⊥ *I* is shown in the inset on the right.

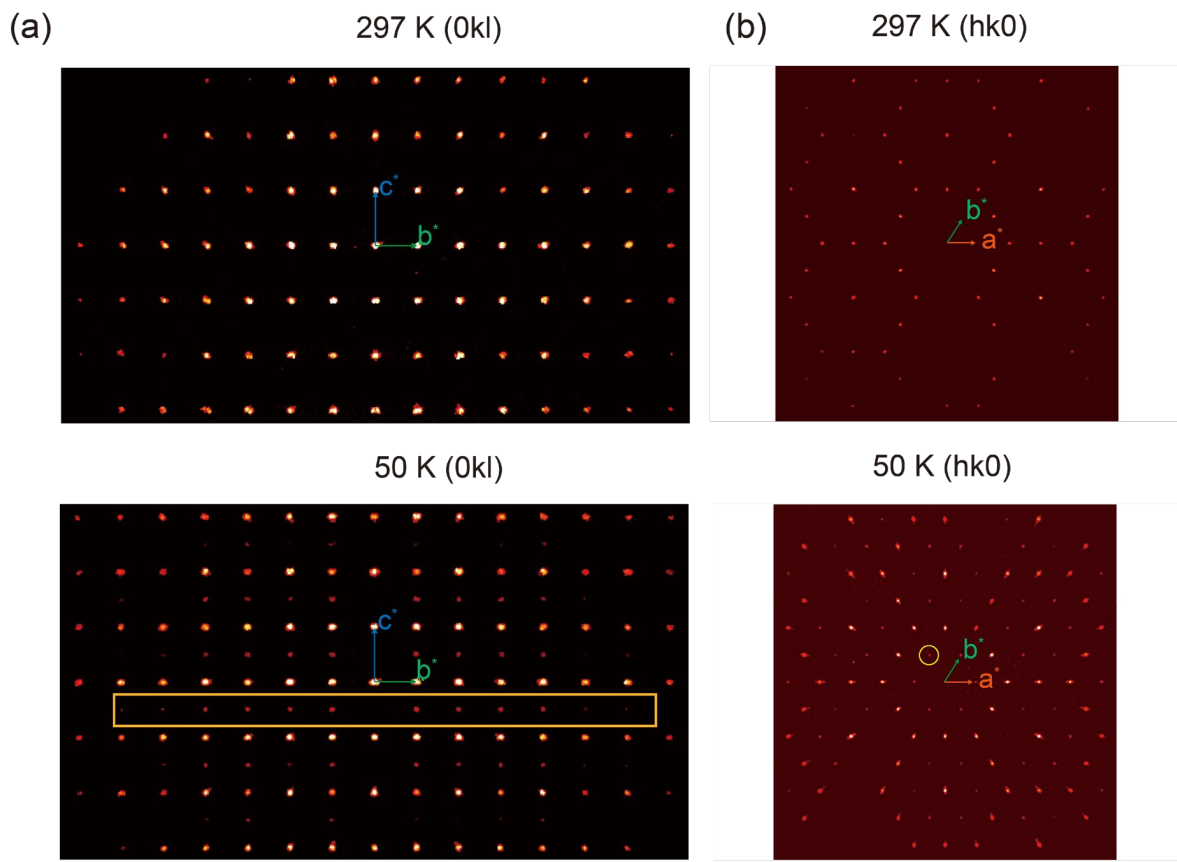

FIG. 2. X-ray diffraction pattern. (**a**) Projection of diffraction pattern along the *a*-axis (0kl plane) at 297 K and 50 K. The appearance of superlattice peaks (marked by the orange rectangle) at 50 K is due to the doubling of *c*-axis. (**b**) Projection of diffraction pattern along the *c*-axis (hk0 plane) at 297 K and 50 K. The new peaks (marked by yellow circle) appearing at 50 K become visible due to the distortion of Co lattice.

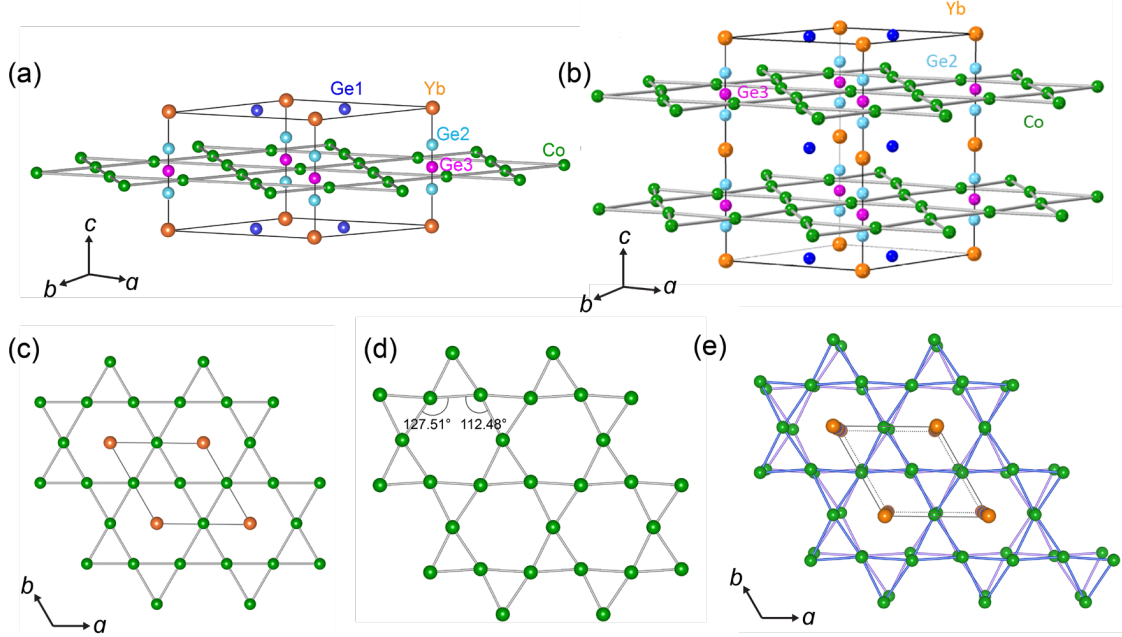

FIG. 3. The crystal structures of Yb$_{0.5}$Co$_3$Ge$_3$ at 297 K and 50 K. (a) and (c) are the crystal structures at 297 K with space group of *P6/mmm* and projection down the c-direction of the Co Kagome network. (b) The crystal structure at 50 K with space group of *P6$_3$/m*. (d) The distortion of one of the Co Kagome layers at 50 K with the modified bond angles of the hexagons labelled. (e) Two layers of the distorted Co Kagome layers unit cell showcasing their rotation relative to each other.

Table 1. Unit Cell Parameters from Single Crystal X-ray Diffraction

|  | 297 K | 50K |
|---|---|---|
| Formula | Yb$_{0.5}$Co$_3$Ge$_3$ | Yb$_{0.5}$Co$_3$Ge$_3$ |
| Space group | *P6/mmm* | *P6$_3$/m* |
| Lattice parameters |  |  |
| a (Å) | 5.0949(10) | 5.0705(4) |
| c (Å) | 3.9136(9) | 7.7780(9) |
| V (Å$^3$) | 87.98(4) | 173.18(3) |
| Z | 1 | 2 |